\documentclass[twocolumn]{aastex63}
\usepackage{bm}
\shorttitle{The pulsar X-ray and gamma-ray emission}
\shortauthors{Cao \& Yang}
\begin{document}
\title{The combined X-ray and $\gamma$-ray modeling of millisecond pulsars PSR J0030+0451 in the dissipative  magnetospheres}
\author{Gang Cao}
\author{Xiongbang Yang}
\affiliation{Department of  Mathematics, Yunnan University of Finance and Economics, Kunming 650221, Yunnan, P. R. China; gcao@ynufe.edu.cn,xbyang@ynufe.edu.cn}
\begin{abstract}
Modeling of the NICER X-ray light curves of  millisecond pulsars PSR J0030+0451 provides a strong evidence for the existence of non-dipole magnetic fields. We study the X-ray and $\gamma$-ray emission of  PSR J0030+0451 in the dissipative dipole plus off-centred quadrupole magnetospheres. The dissipative  FF+AE dipole magnetospheres by combining force-free (FF) and Aristotelian  electrodynamics (AE)  are solved by  a 3D pseudo-spectral method  in the rotating coordinate system.  We use the  FF+AE dipole plus off-centred quadrupole fields with minimum free parameters to reproduce two hotspot configurations found by the NICER observations. The X-ray and  $\gamma$-ray emission from  PSR J0030+0451 are simultaneously computed by using a ray-tracing method and  a particle trajectory method. The modelled X-ray and $\gamma$-ray emission is then directly compared with those of PSR J0030+0451 from the NICER and Fermi observations. Our results can
well reproduce the observed trends of the NICER X-ray and Fermi $\gamma$-ray emission for PSR J0030+0451.
\end{abstract}


\section{Introduction}
Pulsars are thought to be  rapidly rotating and highly magnetized neutron stars. Millisecond pulsars (MSPs) are a class of  the old pulsars with  short rotation periods  in the range of 1 -- 30 \rm ms. They have  low surface magnetic fields with the order of  $B\sim10^{8}$ G and small spin-down rates with the order of  $\dot{P }\sim10^{-20}$ compared to the young pulsars. These objects can produce the multi-wavelength electromagnetic spectra from the  radio up to  $\gamma$-ray bands. The  $\gamma$-ray emission are expected to originate from well above the neutron star surface, the Fermi $\gamma$-ray emission can thus be used to diagnose the structure of the outer magnetosphere. The thermal X-ray emission would originate from the hotspot regions of the neutron star surface, which can provide a excellent probe of surface magnetic field structure. Therefore, the combined X-ray  and $\gamma$-ray modeling can  put the strong constraints on the magnetic field configuration of MSPs.

Recently, Neutron Star Interior Composition Explorer (NICER)  is accurately measuring the  X-ray light emission of MSPs.  One of the first NICER targets is the  MSP PSR J0030+0451 with  the rotation period of $P=4.87 \, \rm ms $, which is also the first $\gamma$-ray MSP  observed by Fermi-LAT \citep{abd09}. The light curves from the NICER observations show double-peak X-ray profiles with $\sim 180^{\circ}$ peak separation. The detailed fitting of the X-ray light curves precisely constrain the mass and radius of the neutron star with the surface hotspot models \citep{ril19,mil19}. They present two  hotspot models with a  circular shape and a crescent shape in the southern hemisphere. Such a configuration provides a strong evidence for the existence of non-dipolar surface magnetic field, because the  dipole fields produces  two near-circular antipodal hotspots. The hotspots are expected to be heated by high-energy particles  from the magnetospheres. The hotspot shapes is determined by the footprints of of the open magnetic field lines, which is related to the structure of  the global pulsar magnetospheres. Therefore, the key to understand the X-ray and $\gamma$-ray emission is to solve the complicated magnetospheric configurations.

Significant advances have been achieved towards the self-consistent modeling of the pulsar magnetospheres. The vacuum magnetosphere is the first solution to describe the global
pulsar magnetosphere \citep{deu55}, which can provide  a reference solution to understand more complicated  magnetospheric structures thanks to the analytic expression. It is well established that the pulsar magnetospheres should be the  plasma-filled magnetospheres with  plasmas from the pair production \citep{gol69}. A zeroth-order approximation FF model is developed to construct the plasma-filled magnetospheres \citep{con99,spi06,kal09,con10,pet12,cao16a,pet16,car18,kim24,dim25,ski25}, which can precisely describe the magnetic field  geometry by including the modification of  plasma currents on the field structure. However, the FF solutions can not accelerate the particles to produce the observed pulsed emission. The dissipative resistive models are then developed to model the pulsar magnetosphere with a prescribed conductivity \citep{li12,kal12,cao16b,pet25}, which can introduce the local accelerating electric field to produce  the observed pulsed emission. However, there are no clear physical motivation for these  arbitrary conductivity in the resistive model. The kinetic particle-in-cell (PIC) methods are later developed to address  the  microscopic physical processes of particle acceleration and creation \citep{phi15,cer16,kal18,bra18,sou24}. However, the PIC method can not use real pulsar parameters to compute the pulsar magnetosphere and predict the pulsar $\gamma$-ray emission due to the large separation between the macroscopic scales to the microscopic scales. Both resistive and PIC simulations produce a near FF magnetosphere with a current sheet outside the light cylinder, and the current sheet is suggested to be the main site of particle acceleration and $\gamma$-ray emission \citep{bai10,kal14,har15,cer16,phi18,cao19,yang21,har21,kal23}. Recently, a new FF+AE model is proposed to construct the pulsar magnetosphere \citep{con16,pet20a,cao20,pet22,cao24a}, which can  produce a near FF  solution with the accelerating electric field  only near the current sheet. Moreover, the FF+AE model can  use real pulsar parameters to predict the pulsar $\gamma$-ray emission.

The multiwavelength emission of PSR J0030+0451 have been studied based on different magnetospheric models \citep{che20,kal21,car23,pet23a}. \citet{pet23a} used a vacuum dipole plus off-centred dipole magnetosphere to  model the radio and $\gamma$-ray  emission of PSR J0030+0451. \citet{car23} used a FF dipole magnetosphere to model the X-ray emission  of PSR J0030+0451, while Fermi $\gamma$-ray emission  is not simultaneously modeled  in their study. \citet{che20} used a FF dipole plus off-centred quadrupole magnetospheres to model the  X-ray and  $\gamma$-ray emission  of PSR J0030+0451, where  a uniform emissivity modulated by the local parallel force-free current is assumed to compute the $\gamma$-ray emission. \citet{kal21} used an off-centred FF dipole plus off-centred quadrupole magnetospheres to model the  X-ray and  $\gamma$-ray of PSR J0030+0451. However, they only used an approximate accelerating electric field scaled with a finite conductivity to  compute $\gamma$-ray emission.  In fact, the $\gamma$-ray emission of PSR J0030+0451 is not self-consistently  computed by using the  accelerating electric fields from  magnetosphere simulation in all these studies. It is shown that  the Fermi observed features of the pulsar  $\gamma$-ray emission can  generally be reproduced by the FF+AE magnetospheres \citep{cao22,cao24b}. Moreover, the FF+AE magnetosphere can also generally reproduced the observed trends of the pulsar multiwavelength emission from the optical to TeV $\gamma$-ray bands \citep{yang24}. Motivated by agreement of the FF+AE magnetospheres with observations, we use a  FF+AE dipole plus  off-centred  quadrupole magnetosphere to explore the  X-ray and $\gamma$-ray emission of PSR J0030+0451, where the $\gamma$-ray emission of PSR J0030+0451 is  self-consistently computed by using the  accelerating electric fields from the FF+AE magnetospheric simulations. In this paper, we use a FF+AE dipole plus off-centred  quadrupole fields to obtain the polar cap shapes similar to  the NICER observation results. A ray-tracing method is developed to compute the thermal X-ray emission, and a particle trajectory method is used to self-consistently compute the $\gamma$-ray emission from curvature radiation based on the  accelerating electric field from  magnetosphere simulation. The predicted  X-ray and $\gamma$-ray emission is then directly compared with the NICER and Fermi observation for PSR J0030+0451.\\

\section{Magnetospheric models}
In our model, the FF+AE dipole magnetospheres by combining the FF and AE method are computed by a 3D spectral code for  a range of dipole inclination angles, the vacuum dipole plus
off-centred quadrupole field is firstly used to find a range of initial dipole and quadrupole parameters with the polar caps similar to the NICER results, these initial dipole and quadrupole parameters are then used to construct a range of the FF+AE dipole plus off-centred quadrupole magnetospheres,  the final parameter solutions for the FF+AE dipole plus off-centred quadrupole magnetospheres are determined by simultaneously fitting the NICER X-ray and Fermi $\gamma$-ray emission.\\
\subsection{Vacuum Dipole and  Off-centred Quadrupole Magnetospheres}
The pulsar magnetospheres are expected to be the plasma-filled magnetospheres instead of the vacuum  magnetospheres. However, the vacuum magnetosphere can be served as a reference solution to explore more complicated  field configuration thanks to the analytic expression. We use the centred dipole plus the off-centred quadrupole in vacuum to find the trial magnetic field configuration, which can reproduce two hotspot shapes similar to the NICER observed ones.

The centred dipole field with magnetic dipole moment $\bm{m}_D=(\sin\chi_{D}\cos\phi_{D},\cos\chi_{D}\sin\phi_{D},\cos\chi_{D})$ depicted by dipole inclination angle $\chi_{D}$ and  azimuth angle $\phi_{D}$  is given by
\begin{equation}
\bm{B}_{\rm D}=B_{\rm D}\frac{R_{*}^3}{r^3}\left[\frac{3(\bm{m}_D\cdot\bm{r})}{r^2}\bm{r}-\bm{m}_D\right],
\end{equation}
Where  $B_{D}$ is surface dipole magnetic field, $R_{*}$ is the neutron star radius.

The aligned off-centred  quadrupole field with magnetic quadrupole moment $\bm{m}_Q=0$ is given by
\begin{equation}
\bm{B}_{\rm Q}=B_{\rm Q}\frac{R_{*}^4}{\left|\bm{r}-\bm{r}_Q\right|^4}\left\{\frac{1}{2}(3\cos^2\theta-1),\cos\theta\sin\theta,0\right\},
\end{equation}
where $B_{\rm Q}=f_{\rm QD}B_{\rm D}$ is surface quadrupole magnetic field, $\bm{r}_Q=(x_{Q},y_{Q},z_{Q})$ is the  position vector of the off-centred quadrupole. For the oblique off-centred quadrupole  field with    $\bm{m}_Q=(\sin\chi_{Q}\cos\phi_{Q},\cos\chi_{Q}\sin\phi_{Q},\cos\chi_{Q})$ depicted by quadrupole inclination angle $\chi_{Q}$ and  azimuth angle $\phi_{Q}$, we rotate the coordinates in the  ($\chi_{Q}$,$\phi_{Q}$) angles to the aligned  off-centred quadrupole field, the oblique  off-centred quadrupole  field  is then obtained by rotating the aligned quadrupole field in the  ($\phi_{Q}$,$\chi_{Q}$)  angles.

\subsection{FF+AE Dipole Magnetospheres}
The FF+AE magnetospheres are described by the time-dependent Maxwell equations in  the rotating coordinate system \citep{mus05,pet20b}
\begin{eqnarray}
{\partial {\bf B}\over \partial t'}=-{\bf \nabla} \times ({\bf E}+{\bm V}_{\rm rot}\times{\bf B}) ,\\
{\partial  {\bf E}\over \partial t'}={\bf \nabla} \times ({\bf B}-{\bm V}_{\rm rot}\times{\bf E})-{\bf J}+{\bm V}_{\rm rot}\nabla\cdot{\bf E},\\
\nabla\cdot{\bf B}=0\;,\\
\nabla\cdot{\bf E}=\rho_{\rm e}\;,
\end{eqnarray}
where  $\rho_{\rm e}$ is the charge density, ${\bm V}_{\rm rot}={\bf \Omega } \times {\bf r}$ is the corotating velocity.
The FF+AE current density ${\bf J }$ is  defined  by introducing the pair multiplicity $\kappa$ \citep{cao20}
\begin{eqnarray}
{\bf J }=  \rho_e  \frac {{\bf E} \times {\bf B}}{B^2+E^2_{0}}+ (1+\kappa)\left|\rho_e\right| \frac{ (B_0{\bf {B}}+E_0{\bf {E}}) }{ B^2+E^2_{0}},
\label{Eq6}
\end{eqnarray}
where  $B_0$ and $E_0$ are the electromagnetic field strength in the frame where ${\bf E}$ and ${\bf B}$ are parallel, and $E_0$  is  the effective accelerating electric component $E_{\rm acc}$.  The quantities $B_0$ and $E_0$ are deduced from the electromagnetic invariants
\begin{eqnarray}
B^2_{0}-E^2_{0}={\bf B}^2-{\bf E}^2, \,\, E_{0}B_{0}={\bf E}\cdot {\bf B},  \,\, E_{0}\geq0.
\end{eqnarray}

\begin{figure}
\center
\begin{tabular}{cccccc}
\includegraphics[width=7.0cm,height=6.cm]{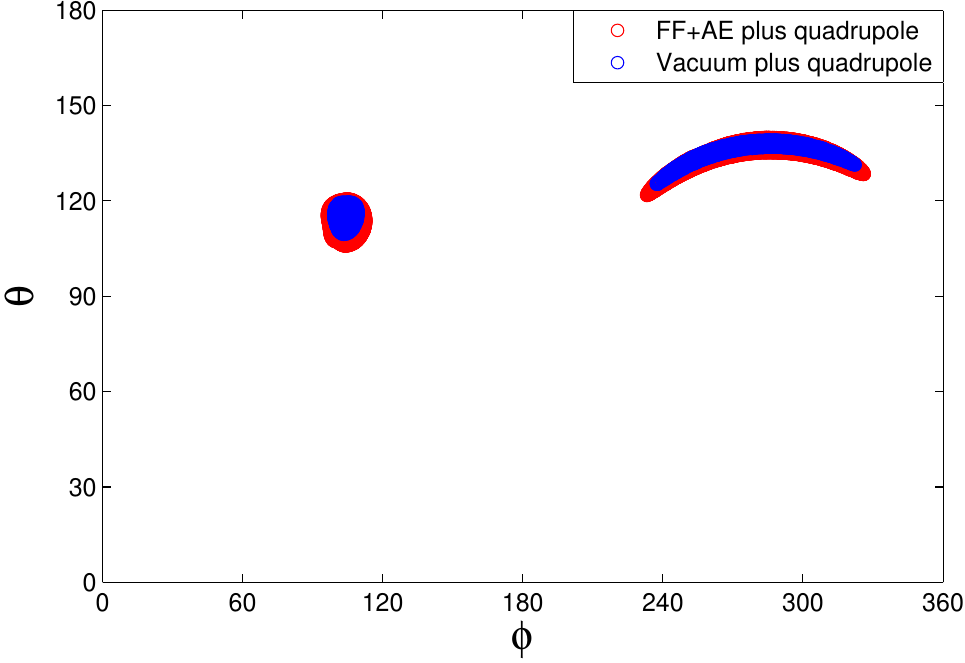}
\end{tabular}
\caption{The polar caps from the FF+AE dipole plus off-centred quadrupole fields (the red curves) and  the vacuum  dipole plus off-centred quadrupole  fields (the  blue curves).  }
\label{Fig1}
\end{figure}

\begin{figure}
\center
\begin{tabular}{cccccc}
\includegraphics[width=7.cm,height=6.5cm]{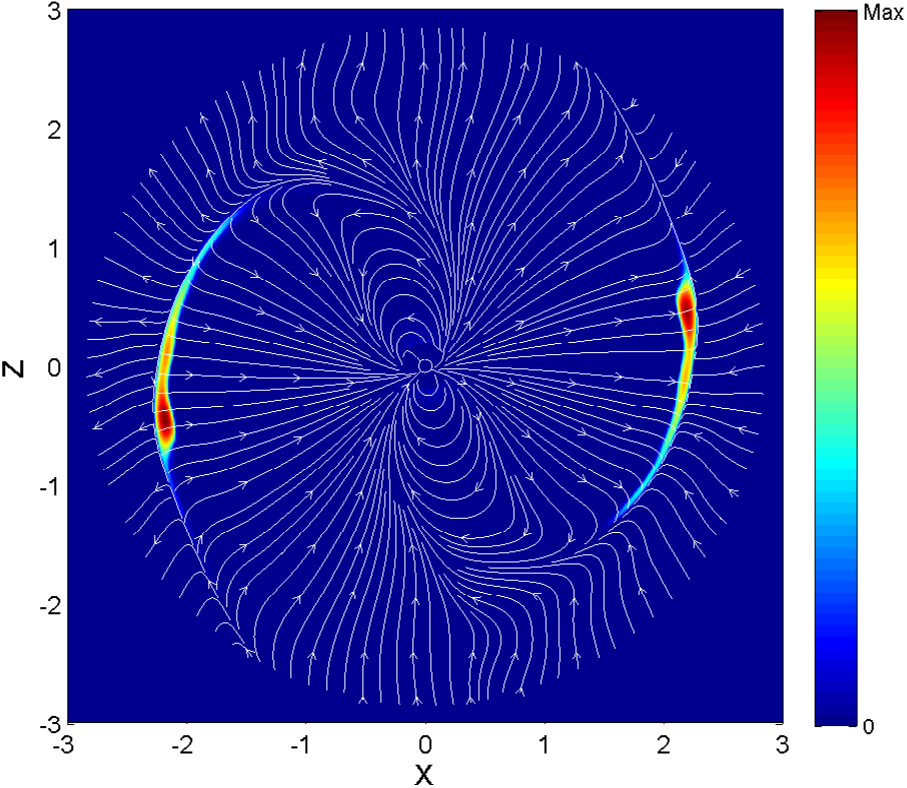}
\end{tabular}
\caption{The Distribution of the magnetic field lines and the accelerating electric fields from the FF+AE dipole plus off-centred quadrupole field.  }
\label{Fig2}
\end{figure}

A 3D spectral method is used to solve the time-dependent Maxwell equations  with the FF+AE  current density in  the rotating coordinate system \citep{cao22}.
The initial magnetic field is set to be an  oblique vacuum dipole  with magnetic inclination angles $\chi_D$ from $0^\circ$ to $90^\circ$ with a $5^\circ$ interval.
A combined three order Runge--Kutta and Adam--Bashforth method is used to advance the electromagnetic field at each time step.
A rotating electric field ${\bf {E}} = -( {\bf \Omega } \times {\bf r} ) \times {\bf B}$ is used to impose the inner boundary condition at the stellar surface.
A outgoing boundary condition is implemented  to prevent inward  reflection   from the artificial outer boundary.
A high pair multiplicity $\kappa=3$ is chosen to produce a near FF solution with the accelerating electric field   in the current sheet outside the light cylinder.
The FF+AE magnetospheres  is constructed by applying the FF description where $E \leq B$ and the AE  description where $E > B$.  A high resolution of $N_r \times N_{\theta} \times N_{\phi}=257 \times 64 \times 128$ is used to obtain the high-precision FF+AE magnetosphere from  the stellar  surface  $r=0.2 $ $R_{\rm L}$ to  $r=3 $ $R_{\rm L}$.
The FF+AE magnetosphere  stabilizes to a near FF solution for a higher $\kappa>3$ value, but the accelerating electric field  $E_{\rm acc}$ decreases with increasing $\kappa$ (see \citealt{cao22}). Therefore, we compute  the $E_{\rm acc}$ value at  $\kappa>3$   by a scaling relation \citep{cao24b}
\begin{eqnarray}
E_{\rm acc}=E_{\rm acc,0}\frac{\kappa_0}{\kappa}\qquad(\kappa>\kappa_0)
\end{eqnarray}
where $E_{\rm acc,0}$ corresponds to the accelerating electric field at $\kappa_0=3$.

\subsection{Magnetospheric Configuration}
There are 8 free parameters  in our magnetospheric  model, they are $(\chi_{D},\phi_{D},\chi_{Q},\phi_{Q},x_{Q},y_{Q},z_{Q}, f_{QD})$. To reduce the free parameters, we place the off-centred quadrupole position to the z-axis with $\bm{r}_Q=(0,0,z_{Q})$, and magnetic dipole moment is placed to the x--z plane with $\phi_{D}=0$. Therefore, the total model parameters are reduced to the 5 free parameters  of $(\chi_{D},\chi_{Q},\phi_{Q},z_{Q},f_{QD})$, which defines  the magnetic field configuration with minimum free parameters. The neutron star surface is randomly discretized  in the $(\theta,\phi)$ directions with a surface resolution of $5000\times5000$, a fourth order Runge--Kutta method is used to find the open magnetic line regions on the stellar surface.

We use the vacuum  dipole plus off-centred quadrupole fields with minimum  parameter assumption to find a range of initial free parameters with polar caps similar to the NICER results.  Figure \ref{Fig1} shows an example of the polar caps from the  vacuum  dipole plus off-centred quadrupole  fields. The model parameters are $\chi_{D}=75^{\circ}$, $\chi_{Q}=53^{\circ}$, $\phi_{Q}=185^{\circ}$, $z_{Q}=-0.05 R_{*}$, $f_{QD}=5$. Our model can produce two  hotspots with a  almost circular shape and a crescent shape in the southern hemisphere, and the two hotspots are  approximately spaced by $\sim 180^{\circ}$ in the $\phi$ direction.  Our  hotspot configurations are similar to  the NICER observed ones.

These initial parameters  are then used to construct the  FF+AE  dipole plus off-centred quadrupole magnetospheres, which can include the modification of plasma charges currents on the polar caps.
Figure \ref{Fig1} also shows  the polar caps from  the   FF+AE dipole plus off-centred quadrupole  fields. It is found that the  polar caps from the  FF+AE dipole plus off-centred quadrupole fields are similar to the   vacuum  dipole plus off-centred quadrupole  ones, but they are more larger and more shifted than the  vacuum  dipole plus off-centred quadrupole ones due to the modification of  magnetospheric plasmas on the polar caps. Figure \ref{Fig2} shows the distribution of magnetic field lines and accelerating electric fields $E_{\rm acc}$ from the  FF+AE dipole plus off-centred quadrupole magnetosphere in the $x$--$z$ plane. The magnetic field lines originate from the two polar caps both in the southern hemisphere. The global field structures are similar to the FF dipole solutions with the current sheet outside the light cylinder, and the accelerating electric fields are only confined in the region near the current sheet.
\begin{figure}
\center
\begin{tabular}{cccccc}
\includegraphics[width=7.0cm,height=5.5cm]{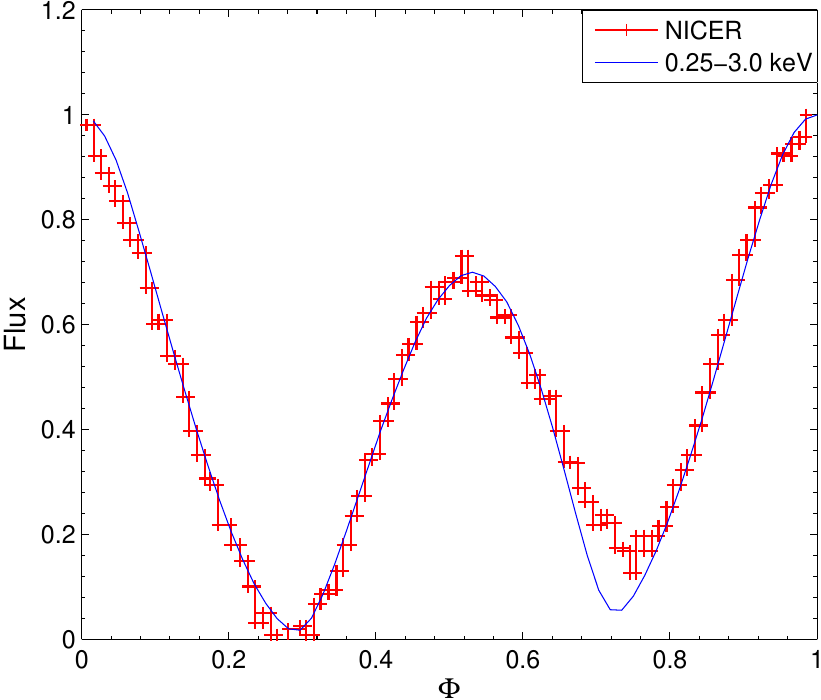}
\end{tabular}
\caption{A  comparison of the predicted X-ray light curves and the NICER observed ones for PSR J0030+0451. The NICER observed data  is taken from \citet{bog19}.  }
\label{Fig3}
\end{figure}

\begin{figure*}
\center
\begin{tabular}{cccccccc}
\\
\includegraphics[width=13 cm,height=6. cm]{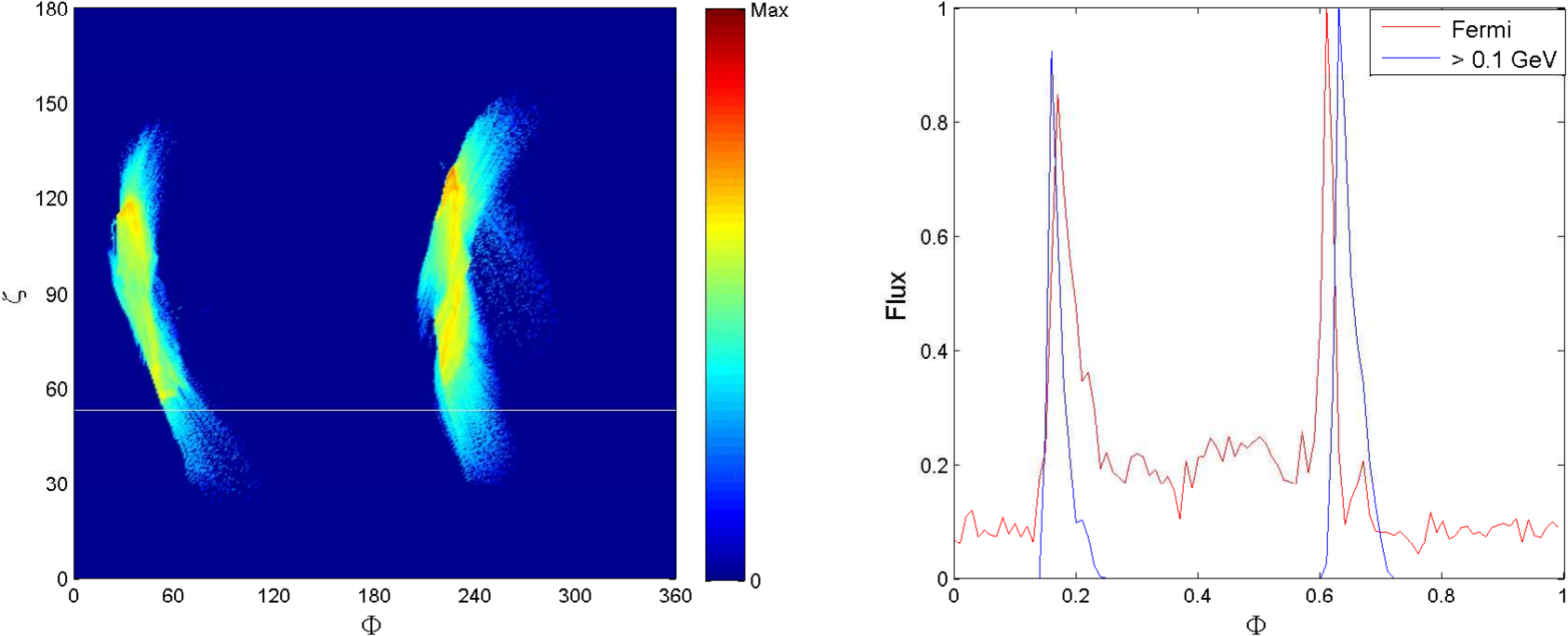}
\end{tabular}
\caption{The sky maps and a  comparison of the predicted $\gamma$-ray light curves and the Fermi observed ones for PSR J0030+0451.  The Fermi observed data is taken from  data taken from \citet{abd13}. \\}
\label{Fig4}
\end{figure*}
\section{Ray-tracing Method for X-rays}
The pulsar X-ray thermal emission can be computed by  performing the ray-tracing of the emitting photons from hot spots  on the  surface of the neutron star. We use a forward ray-tracing method to  trace the photon paths from the emitting region to observer in the spherical Schwarzschild metrics (see \citealt{pou06, bog07}). A surface hot spot is defined  at the co-latitude $\theta$ and azimuthal angle $\phi$ with the unit vector of $\bm{n}=(\sin\theta\, \cos\phi, \sin\theta \, \sin\phi, \cos\theta)$. A distant observer is placed at the $x$--$z$ plane with the unit vector of $\bm{k}=(\sin\zeta, 0, \cos\zeta)$,  where $\zeta$ is  the inclination of the line of sight to the spin axis. The deflection angle between the direction of  the hot spot and the line of sight is given by
\begin{eqnarray}
\cos\psi=\bm{k}\cdot\bm{n}=\cos\zeta\cos\theta+\sin\zeta\sin\theta\cos\phi.
\end{eqnarray}
The relation between the deflection angle $\psi$ and the photon emission angle $\alpha$ is given by \citep{lo13,sal18}
\begin{eqnarray}
\psi=2\frac{\sin\alpha}{\sqrt{1-u}}\int_{0}^{1}\frac{xdx}{\sqrt{x^2q+\cos^2\alpha}},
\end{eqnarray}
where $q=[2-x^2-u(1-x^2)^2/(1-u)]\sin^2\alpha$,$u=R_{\rm s}/R_{*}$, $R_{\rm s}=2GM/c^2$ is the Schwarzschild radius.
The time delay from the light bending is given by \citep{lo13,sal18}
\begin{equation}
\triangle t_{\rm ret}=2\frac{R_{*}}{c}\frac{\sin^2\alpha}{1-u}\int_{0}^{1}\frac{xdx}{\sqrt{x^2q+\cos^2\alpha}\,(1+\sqrt{x^2q+\cos^2\alpha})}
\end{equation}

The observed flux  from the area element $dS$ at the  source distance $D$  is given by \citep{bel02,pou06}
\begin{equation}
F(E)=\sqrt{1-u} \, \delta^3  I^{\prime}(E^{\prime},\alpha^{\prime})\cos\alpha\frac{d\cos\alpha}{d\cos\psi}\frac{dS}{D^2},
\end{equation}
where $\cos\alpha^{\prime}=\delta\cos\alpha$, and $E^{\prime}=E/(\delta\sqrt{1-u})$, $E^{\prime}$ is the photon energy in the co-rotating frame,  $I^{\prime}(E^{\prime},\alpha^{\prime})$  is the specific intensity intensity of the emission in the co-rotating frame,  $\delta$ is the  Doppler factor with
\begin{equation}
\delta=\frac{1}{\Gamma(1-\beta\cos\xi)},
\end{equation}
where $\Gamma=\sqrt{1-\beta^2}$, $\beta=v/c$ is the spot velocity with
\begin{equation}
v=\frac{2\pi R_{*}}{P \sqrt{1-u}}\sin\alpha.
\end{equation}
The angle $\xi$ between the photon emission direction and the spot velocity direction is given by
\begin{equation}
\cos\xi=-\frac{\sin\alpha}{\sin\psi}\sin\zeta\sin\phi,
\end{equation}

The X-ray emission from hot spots is approximated as the blackbody spectrum with the uniform temperature $T^{\prime}$ in the co-rotating frame by
\begin{equation}
I^{\prime}(E^{\prime},\alpha^{\prime})=\frac{2}{c^2h^3}\frac{{E^{\prime}}^3}{e^{E^{\prime}/kT^{\prime}}-1} \mathcal{B}(\alpha^{\prime}),
\end{equation}
where $T^{\prime}=\Gamma \, T$, and an anisotropy beaming distribution is introduced  to approximate the atmosphere model given by
\begin{equation}
\mathcal{B}(\alpha)=\cos^{b}\alpha,
\end{equation}
where $b$ is the anisotropy index, and $b \sim 0.5-1.0 $ is a  good approximation to the  atmosphere model \citep{kal21}. We find that the NICER X-ray profile of PSR J0030+0451 can be well reproduced for $b=0.65$.

The photon emission direction $\bm{k}_0$ is given by
\begin{equation}
\bm{k}_0=\frac{\sin\alpha\bm{k}+\sin(\psi-\alpha)\bm{n}}{\sin\psi}.
\end{equation}
The observed phase is determined by including the rotation and time delay effects
\begin{equation}
\phi_{\rm obs}=\phi_{\rm em}+\Omega \triangle \, t_{\rm ret} - \phi_{\rm rot}
\end{equation}
where $\phi_{\rm em}$ is the phase of the emitting photon, $\phi_{\rm rot}=\Omega \, \triangle t$ is the rotation  phase.

The hot spots are randomly discretized into a series of small area element in the $(\theta,\phi)$ grids. We use a forward ray-tracing method to compute the photon emission direction and the X-ray flux from each small area element. The X-ray light curves is constructed by accumulating all the X-ray flux from each small area element   at a constant view angle $\zeta$.

\section{Particle trajectory method for $\gamma$-rays}
The pulsar $\gamma$-ray emission can be computed by tracing the particle radiation with particle trajectory method in the dissipative magnetosphere. The particle trajectory is defined as the AE velocity formula by \citep{gru12,cai23,pet23b}
\begin{eqnarray}
{\bm v_{\pm}}=  {{\bf E} \times {\bf B}\pm(B_0{\bf {B}}+E_0{\bf {E}}) \over B^2+E^2_{0}},
\label{Eq6}
\end{eqnarray}
where the two signs represent positrons and electrons. They follow different trajectories in the dissipative magnetosphere, which  only depends on  the local electromagnetic field. The Lorentz factor of the emitting particle  is integrated by including the influence of both the local accelerating electric field and the curvature reaction
\begin{eqnarray}
\frac{d\gamma}{dt}=\frac{q_{\rm e}c E_{\rm acc}}{m_{\rm e}c^2}- \frac{2q^2_{\rm e} \gamma^4}{3R^2_{\rm CR}m_{\rm e}c}.
\end{eqnarray}
The  spectrum of the curvature radiation from individual emitting particle with Lorentz factor $\gamma$  is given by
\begin{eqnarray}
F(E_{\gamma},r)=\frac{\sqrt{3} e^2 \gamma}{2 \pi \hbar R_{\rm CR} E_{\gamma}}F(x)\;,
\end{eqnarray}
where $x=E_{\gamma}/E_{\rm cur}$, $E_{\gamma}$ is the curvature photon energy, $E_{\rm cur}=\frac{3}{2}c\hbar\frac{\gamma^3}{R_{\rm CR}}$ is the characteristic curvature photon energy,   the function $F(x)$ is defined by
\begin{equation}
F(x)=x\int_{x}^{\infty}{K_{\rm 5/3}}(\xi)\;d\xi,
\end{equation}
 and the curvature radius $R_{\rm C}$ is computed by
\begin{eqnarray}
R_{\rm CR}=\left|\frac{ d\bm{\beta}}{dl}\right|^{-1}.
\label{Eq6}
\end{eqnarray}
The direction of the photon emission $\bm{\eta}$ is locally tangent to the  direction of particle motion $\bm{\beta}=\bm{v}/c$, the photon emission angles is given by
\begin{eqnarray}
\mu_{\rm em}=\beta_{z}, \quad \phi_{\rm em}=\rm atan \left( \frac{\beta_y}{\beta_x} \right),
\end{eqnarray}
where $\zeta=\rm{acos}(\mu_{\rm em})$. The observed phase is determined  by including the rotation and time-delay effects
\begin{eqnarray}
\phi_{\rm obs}=\phi_{\rm em}+{\bf r_{\rm em}} \cdot {\bm \eta}_{\rm em}/r_{\rm L}-\Omega \triangle t .
\end{eqnarray}

\begin{figure}
\center
\begin{tabular}{cccccc}
\includegraphics[width=7.5cm,height=6.cm]{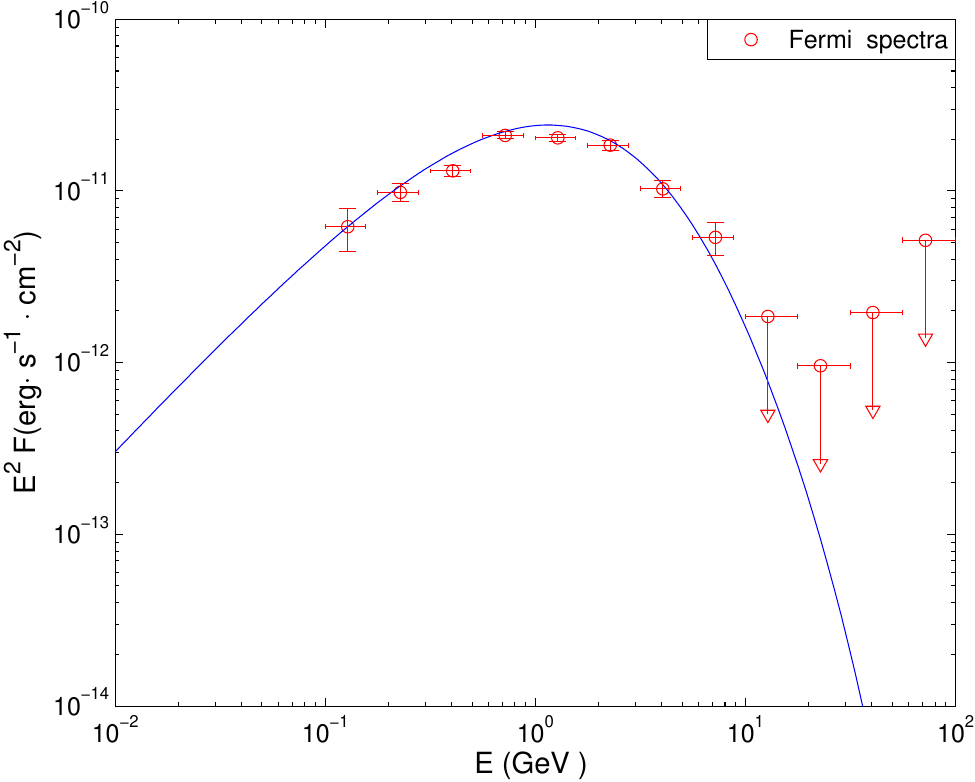}
\end{tabular}
\caption{A  comparison of the predicted $\gamma$-ray phase-averaged spectra and the Fermi observed ones for PSR J0030+0451. The Fermi observed data  is taken from \citet{abd13}.  }
\label{Fig5}
\end{figure}

Positrons and electrons are randomly  injected from  hot spots on the  surface of the neutron star. We use the particle trajectory method to compute the  photon spectrum of curvature radiation  and  the direction of the photon emission  from each emitting particle.  The sky maps are constructed by accumulating all radiation photons  in the ($\zeta$,$\phi$) plane.  The $\gamma$-ray light curves and  spectra is produced by cutting the sky maps at a constant view angle $\zeta$.

\section{results}
The NICER X-ray emission can be used to diagnose the surface magnetic field structures, while the Fermi $\gamma$-ray emission can provide a probe of  the outer magnetic field structures. Therefore, the combined X-ray and $\gamma$-ray modeling can provide  a stronger constraints on the global magnetic field structures by breaking the field degeneracies. We use the polar caps in figure \ref{Fig1} to produce the X-ray light curves of  PSR J0030+0451. The modeling parameters are chosen to be the stellar mass $M_*=1.34 \, M_{\odot}$, the stellar radius $R_{*}=1.27 \times 10^6 \, \rm cm$, the hotspot temperature $T= 1.3 \times 10^6 \, \rm K$ and the observer angle $\zeta=54^{\circ}$, which are same to those from the result of \citet{ril19}. We show the predicted and NICER observed X-ray light curves  in the 0.25$-$3.0 keV bands in Figure \ref{Fig3}. It is find that our results can well reproduce the observed trends of the NICER X-ray light curves. The predicted double-peak X-ray profile with the  phase separation of $180^{\circ}$ is in good agreement with the NICER observed one.

We show the sky maps and the predicted $\gamma$-ray light curves from curvature radiation in the $>0.1 \rm \, GeV$ bands in Figure \ref{Fig4}. The model parameters are  $B_{*}=2.5 \times10^{8} \, \rm G$, $P=4.87 \, \rm ms$ and $\kappa=6$. We see that the emission patten from the sky map shows the characteristic of radiation  with two bright caustics from the current sheet. The peak separation and the peak ratio of the Fermi $\gamma$-ray light curves can  well be reproduced by our model.  It is found that our results can well reproduce the observed trends of the Fermi $\gamma$-ray light curves. We also show the predicted and Fermi observed $\gamma$-ray phase-averaged spectra from curvature radiation  in Figure \ref{Fig5}.  We see that our model can produce an exponential power-law spectrum with the $\sim \rm GeV$ cutoff energy by curvature radiation from the current sheet, and  the predicted phase-averaged spectra provide a good match to the Fermi observed data of PSR J0030+0451. Our results suggest that the curvature radiation from the current sheet is the main radiation mechanism of Fermi $\gamma$-ray emission \citep{kal14,har18,cao24b,ini25}.

\section{Discussion and Conclusions}
We use the dissipative FF+AE  dipole plus off-centred  quadrupole magnetospheres to study the  X-ray and  $\gamma$-ray emission  of PSR J0030+0451. The high-resolution  FF+AE dipole magnetospheres are computed by  a 3D pseudo-spectral method  in the rotating coordinate system. It is find that the   FF+AE dipole  plus off-centred  quadrupole magnetosphere  are similar to the FF solutions with the accelerating electric fields only near current sheet outside the LC. We then use the  FF+AE  dipole plus off-centred  quadrupole fields with  minimum free parameters to obtain the polar cap shapes, which are similar to the hotspot configuration from NICER observations. The thermal X-ray emission from two hotspots is produced by   a develop ray-tracing method,  and the $\gamma$-ray emission from curvature radiation is also simultaneously computed by  a particle trajectory method based on the accelerating electric field from the FF+AE simulation. We then directly compare the modelled  X-ray and  $\gamma$-ray emission with those of PSR J0030+0451 from the NICER and Fermi observations. Our results can well reproduce
the observed trends of the NICER X-ray and Fermi $\gamma$-ray  emission for  PSR J0030+0451.


The  X-ray and  $\gamma$-ray emission  of PSR J0030+0451 was also simultaneously modelled by the  FF dipole plus off-centred quadrupole magnetospheres \citep{che20,kal21}. However, the $\gamma$-ray emission is not self-consistently computed by using the  accelerating electric fields from magnetospheric simulations in these studies. Our model provides a more self-consistent modeling of the $\gamma$-ray emission by using the  accelerating electric fields from the FF+AE simulations, which can both well reproduce the NICER X-ray and Fermi $\gamma$-ray emission of PSR J0030+0451. Our combined X-ray and $\gamma$-ray modeling obtain a dipole inclination angle of $75^{\circ}$, which is consistent with the result from  the combined radio and $\gamma$-ray modeling in \citet{pet23a}. A dipole inclination angle of $\sim80^{\circ}$ is also found from the combined X-ray and $\gamma$-ray modeling in \citet{che20} and \citet{kal21}. Therefore, we suggest a dipole inclination angle of $70^{\circ}<\chi_D<90^{\circ}$ as a robust estimates for PSR J0030+0451. A uniform temperature distribution from the hotspot regions is assumed in our X-ray modeling. We will include a non-uniform temperature distribution from the hotspot regions to model the  X-ray emission of PSR J0030+0451 in the next step. A growing number of MSPs are expected to detected by  NICER with high-precision   measurement of the X-ray light curves. We will further explore the general properties of our model for a group of MSPs with the X-ray and $\gamma$-ray observations. The X-ray emission from MSPs may show significant polarization properties in the present of the multipolar fields. Future X-ray polarization measurement can provide an additional dimension  to break the parameter degeneracy and constrain the surface magnetic field structure. We will extend our model to present the predictions of  MSPs multi-wavelength radiation and polarization in the future work.

\acknowledgments
We thank the anonymous referee for valuable comments and suggestions.
We would like to thank  J\'{e}r$\hat{\rm o}$me P\'{e}tri and Joaquin Pelle for some useful discussions.  We acknowledge the financial support from the National Natural Science Foundation of China 12003026, 12373045 and 12403057, and the Basic research Program of Yunnan Province 202001AU070070 and 202301AU070082.



\begin{thebibliography}{}
\bibitem[Abdo et al. (2019)]{abd09} Abdo, A. A., Ackermann, M., Atwood, W. B., et al. 2009, ApJ, 699, 1171
\bibitem[Abdo et al. (2013)]{abd13} Abdo, A. A., Ajello, M., Allafort, A., et al. 2013, ApJS, 208, 17
\bibitem[Bai \& Spitkovsky (2010)]{bai10} Bai, X.N., \& Spitkovsky, A. 2010, ApJ, 715, 1282
\bibitem[Beloborodov (2002)]{bel02} Beloborodov, A. M. 2002, ApJ, 566, L85
\bibitem[Brambilla et al. (2018)]{bra18} Brambilla, G., Kalapotharakos, C., Timokhin, A. N., Harding, A. K. \& Kazanas, D. 2018, ApJ, 858, 81
\bibitem[Bogdanov et al. (2007)]{bog07} Bogdanov, S., Rybicki, G. B., \& Grindlay, J. E. 2007, ApJ, 670, 668
\bibitem[Bogdanov et al. (2019)]{bog19} Bogdanov, S., Guillot, S., Ray, P. S., et al. 2019, ApJL, 887, L25
\bibitem[Cao et al. (2016a)]{cao16a} Cao, G., Zhang, L., \& Sun, S. N. 2016a, MNRAS, 455, 4267.
\bibitem[Cao et al. (2016b)]{cao16b} Cao, G., Zhang, L., \& Sun, S. N. 2016b, MNRAS, 461, 1068.
\bibitem[Cao \& Yang (2019)]{cao19} Cao, G., \& Yang, X. B. 2019, ApJ, 874, 166
\bibitem[Cao \& Yang (2020)]{cao20} Cao, G., \& Yang, X. B. 2020, ApJ, 889, 29
\bibitem[Cao \& Yang (2022)]{cao22} Cao, G., \& Yang, X. B, 2022, ApJ, 925, 130
\bibitem[Cao et al (2024a)]{cao24a} Cao, G.,  Yang, X. B., \& Zhang, L.  2024a, Universe, 10, 130
\bibitem[Cao \& Yang (2024b)]{cao24b} Cao, G., \& Yang, X. B, 2024b, ApJ, 962, 184

\bibitem[Carrasco et al. (2018)]{car18} Carrasco, F., Palenzuela, C. \& Reula, O, 2018, Phys. Rev. D, 98, 023010
\bibitem[Carrasco et al. (2023)]{car23} Carrasco, F., Pelle, J., Reula, O., Vigan\'{o}, D., \& Palenzuela, C.. 2023, MNRAS, 520, 3151
\bibitem[Cai et al. (2023)]{cai23} Cai, Y., Gralla, S. E., \& Paschalidis, V. 2023, PhRvD, 108, 063018
\bibitem[Chen et al. (2020)]{che20} Chen, A. Y., Yuan, Y., \& Vasilopoulos, G. 2020, ApJ, 893, L38
\bibitem[Contopoulos et al. (1999)]{con99} Contopoulos, I., Kazanas, D., \& Fendt, C. 1999, ApJ, 511, 351
\bibitem[Contopoulos \& Kalapotharakos (2010)]{con10} Contopoulos, I., \& Kalapotharakos, C. 2010, MNRAS, 404, 767
\bibitem[Contopoulos et al. (2016)]{con16} Contopoulos I. 2016, \mnras, 463, L94
\bibitem[Cerutti et al. (2016)]{cer16} Cerutti B., Philippov A. A., \& Spitkovsky, A. 2016, MNRAS, 457, 2401
\bibitem[Deutsch (1955)]{deu55} Deutsch, A. J. 1955, Ann. Astrophys, 18, 1
\bibitem[Dimitropoulos \& Contopoulos (2025)]{dim25} Dimitropoulos, I. \& Contopoulos, I. 2024, arXiv:2410.10716
\bibitem[Goldreich \& Julian (1969)]{gol69} Goldreich, P., \& Julian, W. H. 1969, ApJ, 157, 869
\bibitem[Gruzinov (2012)]{gru12} Gruzinov, A. 2012, arXiv: 1205.3367


\bibitem[Harding \& Kalapotharakos (2015)]{har15} Harding, A. K., \& Kalapotharakos, C. 2015, ApJ, 811, 63
\bibitem[Harding et al. (2018)]{har18}   Harding, A. K., Kalapotharakos, C., Barnard, M., \& Venter, C. 2018, ApJL, 869, L18
\bibitem[Harding et al. (2021)]{har21}    Harding, A. K., Venter, C., \& Kalapotharakos, C. 2021, ApJ, 923, 194
\bibitem[{\'I}{\~n}iguez-Pascual et al. (2025)]{ini25} {\'I}{\~n}iguez-Pascual, D., Torres, D. F., \& Vigan\'{o}, D., 2025, MNRAS, 541, 806
\bibitem[Kalapotharakos \& Contopoulos (2009)]{kal09} Kalapotharakos, C., \& Contopoulos, I. 2009, A\&A, 496, 495
\bibitem[Kalapotharakos et al. (2012)]{kal12} Kalapotharakos, C., Kazanas D., Harding, A., \& Contopoulos, I. 2012, ApJ, 749, 2
\bibitem[Kalapotharakos et al. (2014)]{kal14} Kalapotharakos, C., Harding, A. K., \& Kazanas, D. 2014, ApJ, 793, 97
\bibitem[Kalapotharakos et al. (2018)]{kal18} Kalapotharakos, C., Brambilla, G., Timokhin, A., Harding, A. K., \& Kazanas, D. 2018, ApJ, 857, 44
\bibitem[Kalapotharakos et al. (2021)]{kal21} Kalapotharakos, C., Wadiasingh, Z., Harding, A. K., \& Kazanas, D. 2021, ApJ, 907, 63
\bibitem[Kalapotharakos et al. (2023)]{kal23} Kalapotharakos, C.,  Wadiasingh, Z., Harding, A. K., \& Kazanas, D., 2023, ApJ, 954, 204
\bibitem[Kim et al. (2024)]{kim24} Kim, Y., Most, E. R., Throwe, W., et al. 2024, PhRvD, 109, 123019
\bibitem[Li et al. (2012)]{li12} Li, J., Spitkovsky, A., \& Tchekhovskoy, A. 2012, ApJ, 746, 60
\bibitem[Lo et al. (2013)]{lo13} Lo, K. H., Miller, M. C., Bhattacharyya, S., \& Lamb, F. K. 2013, ApJ, 776, 19
\bibitem[Muslimov \& Harding (2005)]{mus05} Muslimov, A. G., \& Harding, A. K. 2005, ApJ, 630, 454
\bibitem[Miller et al. (2012)]{mil19} Miller, M. C., Lamb, F. K., Dittmann, A. J., et al. 2019, ApJ, 887, L24



\bibitem[P\'{e}tri (2012)]{pet12} P\'{e}tri, J. 2012, MNRAS, 424, 605
\bibitem[P\'{e}tri (2016)]{pet16} P\'{e}tri, J. 2016, MNRAS, 455, 3779
\bibitem[P\'{e}tri (2020a)]{pet20a} P\'{e}tri, J. 2020a, MNRAS, 491, 46
\bibitem[P\'{e}tri (2020b)]{pet20b} P\'{e}tri, J. 2020b, Universe, 6, 15
\bibitem[P\'{e}tri (2022)]{pet22} P\'{e}tri, J. 2022, MNRAS, 512, 2854
\bibitem[P\'{e}tri et al (2023a)]{pet23a} P\'{e}tri, J., Guillot, S., Guillemot, L., et al.  2023a, A\&A, 680, 93
\bibitem[P\'{e}tri (2023b)]{pet23b} P\'{e}tri, J. 2023b, A\&A, 677, 72
\bibitem[P\'{e}tri (2025)]{pet25} P\'{e}tri, J. 2025, A\&A in press, arXiv:2104.09802
\bibitem[Philippov et al. (2015)]{phi15} Philippov, A. A., Spitkovsky, A., \& Cerutti, B. 2015, ApJ, 801, L19
\bibitem[Philippov \& Spitkovsky (2018)]{phi18} Philippov, A. A. \& Spitkovsky, A., 2018, ApJ, 855, 94
\bibitem[Poutanen \& Beloborodov (2006)]{pou06} Poutanen, J., \& Beloborodov, A. M. 2006, MNRAS, 373, 836
\bibitem[Riley et al. (2019)]{ril19} Riley, T. E., Watts, A. L., Bogdanov, S., et al. 2019, ApJ, 887, L21
\bibitem[Salmi et al. (2018)]{sal18} Salmi, T., N\"{a}ttil\"{a}, J., \& Poutanen, J, 2018, A\&A, 618, A161
\bibitem[Skiathas et al. (2025)]{ski25} Skiathas, D., Kalapotharakos, C., Wadiasingh, Z., et al. 2025, ApJ, 994, 131
\bibitem[Spitkovsky (2006)]{spi06} Spitkovsky, A. 2006, ApJ, 648, L51
\bibitem[Soudais (2024)]{sou24} Soudais, S., Cerutti B., \& Contopoulos, I. 2024, A\&A, 690, A170
\bibitem[Yang \& Cao (2021)]{yang21} Yang, X. B., \& Cao, G. 2021, ApJ, 909, 88
\bibitem[Yang \& Cao (2024)]{yang24} Yang, X. B., \& Cao, G. 2024, ApJ, 964, 72

\end{thebibliography}
\end{document}